\documentclass[notitlepage]{revtex4-1}

\usepackage[utf8]{inputenc}
\usepackage[colorlinks=true,citecolor=blue,linkcolor=blue]{hyperref}
\usepackage[normalem]{ulem}
\usepackage{url}
\usepackage{graphicx,wrapfig,float,slashed,cancel}
\usepackage{amsmath,amssymb,epsfig,graphicx,xcolor,stmaryrd}
\usepackage{bm}
\usepackage{enumitem}

\definecolor{darkblue}{RGB}{1, 90, 173}

\begin{document}


\title{Investigation of $P_{cs}(4459)^0$ pentaquark via its strong decay to $\Lambda J/\psi$}

\author{K.~Azizi}
\email{ kazem.azizi@ut.ac.ir}
\thanks{Corresponding author}
\affiliation{Department of Physics, University of Tehran, North Karegar Avenue, Tehran
14395-547, Iran}
\affiliation{Department of Physics, Do\u gu\c s University,
Ac{\i}badem-Kad{\i}k\"oy, 34722 Istanbul, Turkey}
\affiliation{School of Particles and Accelerators, Institute for Research in Fundamental Sciences (IPM) P.O. Box 19395-5531, Tehran, Iran}
\author{Y.~Sarac}
\email{yasemin.sarac@atilim.edu.tr}
\affiliation{Electrical and Electronics Engineering Department,
Atilim University, 06836 Ankara, Turkey}
\author{H.~Sundu}
\email{ hayriye.sundu@kocaeli.edu.tr}
\affiliation{Department of Physics, Kocaeli University, 41380 Izmit, Turkey}

\date{\today}

\preprint{}

\begin{abstract}

Recently the observation of a new pentaquark state, the hidden-charmed strange $P_{cs}(4459)^0$, was reported by the LHCb Collaboration. The spin-parity quantum numbers of this state were not determined as a result of insufficient statistics. To shed light on its quantum numbers, we investigate its decay, $P_{cs}(4459)^0 \rightarrow J/\psi \Lambda $, the mode that this state has been observed, within the QCD sum rule framework.  We obtain the width of this decay  assigning  the  spin-parity quantum numbers of $P_{cs}(4459)^0$ state as $J^P=\frac{1}{2}^-$ and its substructure as diquark-diquark-antiquark. To this end, we first calculate the strong coupling constants defining the considered decay and then use them in the width calculations. The obtained width is consistent  with the experimental observation, confirming the quantum numbers $J^P=\frac{1}{2}^-$ and compact pentaquark nature for $P_{cs}(4459)^0$ state.

\end{abstract}


\maketitle

\renewcommand{\thefootnote}{\#\arabic{footnote}}
\setcounter{footnote}{0}
\section{\label{sec:level1}Introduction}\label{intro}

In the past two decades, starting with the observation of the $X(3872)$~\cite{Choi:2003ue}, we witnessed the observations of  many exotic hadrons candidates for tetraquarks~\cite{Zyla:2020zbs}  and pentaquarks~\cite{Aaij:2015tga,Aaij:2016ymb,Aaij:2019vzc}. The first observation of pentaquark states was announced in 2015 by the LHCb collaboration~\cite{Aaij:2015tga} and two pentaquark states in $J/\psi p$ invariant mass spectrum of the $\Lambda_b^0 \rightarrow J/\psi p K^-$ decays were reported with the following resonance parameters~\cite{Aaij:2015tga}: $m_{P_c(4380)^+}=4380 \pm8 \pm 29~\mathrm{MeV}$, $\Gamma_{P_c(4380)^+}=205 \pm 18 \pm 86~\mathrm{MeV}$ and $m_{P_c(4450)^+}=4449.8 \pm 1.7 \pm 2.5~\mathrm{MeV}$, $\Gamma_{P_c(4450)^+}= 39 \pm 5 \pm 19~\mathrm{MeV}$. The LHCb Collaboration supported this observation later, in 2016, with a full amplitude analysis for $\Lambda_b^0 \rightarrow J/\psi p \pi^-$ decays~\cite{Aaij:2016ymb}. In 2019, a new pentaquark resonance, $P_c(4312)^+$, was reported by the LHCb Collaboration with the following mass and width~\cite{Aaij:2019vzc}: $m_{P_c(4312)^+}=4311.9 \pm 0.7^{ +6.8}_{-0.6}~\mathrm{MeV}$ and $\Gamma_{P_c(4312)^+}=9.8 \pm 2.7 ^{ +3.7}_{-4.5}~\mathrm{MeV}$. Together with the $P_c(4312)^+$ state, the LHCb also announced the split of the peak corresponding to $P_c(4450)^-$ into two peaks which have the following masses and widths: $m_{P_c(4440)^+}=4440.3 \pm 1.3 ^{+ 4.1}_{-4.7}~\mathrm{MeV}$, $\Gamma_{P_c(4440)^+}= 20.6 \pm 4.9^{+8.7}_{-10.1}~\mathrm{MeV}$ and $m_{P_c(4457)^+}=4457.3 \pm 0.6 ^{+ 4.1}_{-1.7}~\mathrm{MeV}$, $\Gamma_{P_c(4457)^+}= 6.4 \pm 2.0^{+5.7}_{-1.9}~\mathrm{MeV}$~\cite{Aaij:2019vzc}. These observations and the advances in experimental facilities and techniques indicate the possibility to observe more exotic states in the future.

On the other hand, there is still uncertainties in the sub-structures and quantum numbers of these observed pentaquark states. In that matter, there are different proposals and theoretical works about these resonances in the literature analyzing their parameters and giving consistent predictions with their observed properties. It is obvious that deeper investigations are required not only  to differentiate these proposals but also to help better identify the nature of these states. Understanding the inner structures and properties of these exotic states may also support their future investigations. Besides, they may provide improvements in understanding the dynamics of the quantum chromodynamics (QCD) in its nonperturbative domain. With their non-conventional quark substructures that are different from the conventional baryons composed of three quarks/antiquarks or mesons composed of a quark and an antiquark, they provide an attractive ground for the understanding of the nonperturbative nature of strong interaction.  Although the investigations of such exotic states extend before their observations, with their observations the pentaquarks have become a hot topic in all these respects.  With these motivations and the excitement brought by their observations, their various properties were investigated widely with different approaches to shed light on their nonspecific sub-structures and quantum numbers. Based on their close masses to  meson-baryon threshold, they were assigned as meson baryon molecular states in Refs.~\cite{Chen:2015loa,Chen:2015moa,He:2015cea,Meissner:2015mza,Roca:2015dva,Azizi:2016dhy,Azizi:2018bdv,Azizi:2020ogm,Chen:2020opr}. They were interpreted with diquark-diquark-antiquark~\cite{Lebed:2015tna,Li:2015gta,Maiani:2015vwa,Anisovich:2015cia,Wang:2015ava,Wang:2015epa,Wang:2015ixb,Ghosh:2015ksa,Wang:2015wsa,Zhang:2017mmw,Wang:2019got,Wang:2020rdh,Ali:2020vee,Wang:2016dzu} and diquark-triquark~\cite{Wang:2016dzu,Zhu:2015bba} models. To investigate their properties in Ref.~\cite{Liu:2017xzo}  a variant of the D4-D8 model and in Ref.~\cite{Scoccola:2015nia} the topological soliton model were used. They were also explained as kinematical effects~\cite{Guo1,Guo2,Mikhasenko:2015vca,Liu1,Bayar:2016ftu}. Besides the observed ones, the possible other candidate pentaquark states were also considered in the literature with different quark contents~\cite{Liu:2020cmw,Chen:2015sxa,Feijoo:2015kts,Lu:2016roh,Irie:2017qai,Chen:2016ryt,Zhang:2020cdi,Paryev:2020jkp,Gutsche:2019mkg,Azizi:2017bgs,Cao:2019gqo,Azizi:2018dva,Zhang:2020vpz,Wang:2020bjt,Xie:2020ckr}.

Recently, in a talk, implications of LHCb measurements and future prospects, the evidence for a pentaquark including a strange quark in its quark content was first announced by the LHCb Collaboration~\cite{Wang111} and later it was reported in the Ref.~\cite{Aaij:2020gdg}. The $P_{cs}(4459)^0$ was observed in $\Xi_b^-\rightarrow J/\psi K^-\Lambda$ decays with the following measured mass and width~\cite{Aaij:2020gdg}:
\begin{eqnarray}
M=4458.8 \pm 2.9^{+4.7}_{-1.1}~\mathrm{MeV},~~~~~~~~~~~~\Gamma = 17.3 \pm 6.5^{+8.0}_{-5.7}~\mathrm{MeV},
\end{eqnarray}
with statistical significance exceeding $3 \sigma$ and there is no determination for its spin parity quantum numbers, yet. With a mass just below $\Sigma_c\bar{D}^*$ threshold, the $P_{cs}(4459)^0$ was interpreted as $\bar{D}^*\Sigma_c$ hadronic molecular state in Ref.~\cite{Chen:2020uif}. The analyses were conducted using QCD sum rule method and the results supported its possibility to be $\bar{D}^*\Sigma_c$ molecular state with either $J^P=\frac{1}{2}^-$ or $J^P=\frac{3}{2}^-$ giving mass values consistent with the experimentally reported one~\cite{Chen:2020uif}. Molecular explanation for the $P_{cs}(4459)^0$ was also discussed in Ref.~\cite{Peng:2020hql} using effective field formalism and the masses were predicted considering the $\bar{D}^{*}\Xi_c$ molecular pictures with $J^P=\frac{1}{2}$ and $J=\frac{3}{2}$ as $4469$~MeV and $4453-4463$~MeV, respectively. With these results the spin of the  $P_{cs}(4459)^0$ state was suggested to possibly be $J=\frac{3}{2}$. In Ref.~\cite{Chen:2020kco} molecular interpretation was taken into account using one-boson-exchange model and $P_{cs}(4459)^0$ was interpreted as a coupled $\Xi_c\bar{D}^{*}/\Xi_c^{*}\bar{D}/\Xi_c^{'}\bar{D}^{*}/\Xi_c^{*}\bar{D}^{*}$ bound state that has $I(J^P)=0(\frac{3}{2}^-)$. In Ref.~\cite{Wang:2020eep} the mass analysis was made via QCD sum rule approach for a pentaquark state containing strange quark with an interpolating current in the scalar-diquark-scalar-diquark-antiquark form. Based on the mass value obtained for the state as $M=4.47\pm 0.11$~MeV, which was consistent with the experimentally observed one, $P_{cs}(4459)^0$ was assigned to have the quantum numbers  $J^P=\frac{1}{2}^-$.

As is seen, the quantum numbers for the $P_{cs}(4459)^0$ state were not determined by the experiment, and from different studies there are different assumptions for its quantum numbers and sub-structure, indicating the necessity for further investigations of the properties of this state. Inspired by this, we investigate the $P_{cs}(4459)^0$ state through its strong decay via QCD sum rule method~\cite{Shifman:1978bx,Shifman:1978by,Ioffe81}. This method has a wide range of applications in the literature, which resulted in successful predictions consistent with the experimental observations. To analyze the pentaquark states within the QCD sum rule approach a proper choice of the interpolating current is necessary. So far, it has been observed that, in various QCD sum rules analyses for observed pentaquark states, the different choices of the interpolating fields, either in the molecular form or in the diquark-diquark antiquark form, were applied. These analyses have resulted in mass predictions that are consistent with the experimental observations. In Refs.~\cite{Wang:2021itn,Wang:2020rdh}, it was pointed out that the hadronic dressing mechanism, which also works for $X$, $Y$, $Z$ states~\cite{Wang:2019iaa},  may compromise the interpretation of these states as a molecule or diquark-diquark-antiquark states considering the result of the QCD sum rules. This may be attributed to the possibility that the pentaquark states may have both the diquark-diquark-antiquark and meson-baryon type Fock components. The pentaquark states may have a typical size of the conventional baryon with a diquark-diquark-antiquark type kernel, and the strong couplings to the meson-baryon pairs may cause  spending a considerable part of their life time as meson-baryon molecules. The local interpolating current in the diquark-diquark-antiquark form can be formed from special superposition of the meson-baryon type interpolating currents and the opposite is also possible, that is a meson baryon current can also be written as a special superposition of diquark-diquark-antiquark type currents that carries the net effect, (see for instance Ref.~\cite{Wang:2019got}). To interpolate the pentaquark state, one can choose either diquark-diquark-antiquark or molecular type currents. Any current type with the same quark structure and quantum numbers of the pentaquark's Fock states may couple to the pentaquark. The main component of the Fock states may give the sub-structure of the pentaquark. For more details, we refer to the Refs.~\cite{Wang:2021itn,Wang:2020rdh,Wang:2018waa,Wang:2019iaa,Wang:2019hyc}. Therefore, besides mass predictions, the investigations of their decay mechanisms, using different choices for interpolating currents, may help to distinguish their substructure with comparisons of the results to experimental results. In this work, to provide the width, we first calculate the strong coupling constants defining the decay $P_{cs}(4459)^0 \rightarrow J/\psi \Lambda $ using three-point QCD sum rule approach with an interpolating current in the scalar-diquark-scalar-diquark-antiquark form of $J^P=\frac{1}{2}^-$. Then,  the obtained results for the strong coupling constants are used to determine the corresponding width value.  We compare the obtained result  with the experimental observation to shed light on the quantum numbers and quark sub-structure of the considered state. 

The organization of the paper is as follows: In  next section we give the details of the QCD sum rule calculations for the strong coupling constants defining the $P_{cs}(4459)^0 \rightarrow J/\psi \Lambda $ decay. The numerical analyses of the obtained sum rules as well as the width of the considered decay are also presented in Sec.~\ref{II}. Last section is devoted to a summary and comparison of the obtained result for the width to that of the experiment.

\section{The strong decay $P_{cs}(4459)^0 \rightarrow J/\psi \Lambda $}\label{II}

In this section the details of the calculations for the strong coupling constants and the width of the strong decay $P_{cs}(4459)^0 \rightarrow J/\psi \Lambda $ and their numerical analyses are given. The correlation function required for the calculations has the following form:
\begin{equation}
\Pi_{\mu} (p, q)=i^2\int d^{4}xe^{-ip\cdot
x}\int d^{4}ye^{ip'\cdot y}\langle 0|\mathcal{T} \{\eta^{\Lambda}(y)
\eta_{\mu}^{J/\psi}(0)\bar{\eta}^{P_{cs}}(x)\}|0\rangle,
\label{eq:CorrF1Pc}
\end{equation}
where the  $\eta^{P_{cs}}$, $\eta^{\Lambda}$ and $\eta^{J/\psi}$ are the interpolating currents of the considered states which have the same quantum numbers with these states and $\mathcal{T}$ is used to represent the time ordering operator. The interpolating currents are given as:
\begin{eqnarray}
\eta^{P_{cs}}&=&\epsilon^{ila}\epsilon^{ijk}\epsilon^{lmn}u^{T}_{j}C\gamma_5 d_k s^{T}_m C\gamma_5 c_{n} C \bar{c}^{T}_a,\nonumber\\
\eta^{\Lambda}&=&\frac{1}{\sqrt{6}}\epsilon^{lmn}\sum_{i=1}^{2}\Big[2(u^{T}_l CA_1^i d_m)A_2^i s_n+(u^{T}_l CA_1^i s_m)A_2^i d_n+(d^{T}_n CA_1^i s_m)A_2^i u_l\Big],\nonumber\\
\eta_{\mu}^{J/\psi}&=&\bar{c}_li\gamma_{\mu}c_l,
\end{eqnarray}
 where the sub-indices, $a,~i,~j,~k,~l,~m,~n$ represent the color indices $u,~d,~s,~c$ are the quark fields, $C$ is charge conjugation operator; and $A_1^1=I$, $A_1^2=A_2^1=\gamma_5$ and $A_2^2=\beta$ is an arbitrary parameter to be determined from the analyses. The above correlation function is calculated in two representations which are called hadronic and QCD representations. The QCD sum rules for the physical quantities are obtained from the matches of the coefficients of the same Lorentz structures attained on both sides. 

In the hadronic representation of the correlation function, the interpolating currents are treated as operators creating or annihilating the hadronic states. To proceed in the calculation of this side, complete sets of related hadronic states that have the same quantum numbers with the given interpolating currents are inserted inside the correlator. After taking four integrals the results turn into 
\begin{eqnarray}
\Pi _{\mu }^{\mathrm{Had}}(p, q)=\frac{\langle 0|\eta^{\Lambda }|\Lambda(p',s')\rangle \langle 0|\eta_{\mu }^{J/\psi}|J/\psi(q)\rangle \langle J/\psi(q) \Lambda(p',s')|P_{cs}(p,s)\rangle \langle P_{cs}(p,s)|\eta^{P_{cs}}|0\rangle }{(m_\Lambda^2-p'^2)(m_{J/\psi}^2-q^2)(m_{P_{cs}}^2-p^2)}+\cdots,  \label{eq:CorrF1Phys}
\end{eqnarray}
where $\cdots$ is used to represent the contributions of higher states and continuum, the $p$, $p'$ and $q$ are the momenta of the $P_{cs}$ and $\Lambda$ and $J/\psi$ states, respectively. The matrix elements in this result are defined in terms of the masses and current coupling constants, and they have the following forms:
\begin{eqnarray}
\langle 0|\eta^{P_{cs}}|P_{cs}(p,s)\rangle &=&\lambda_{P_{cs}} u_{P_{cs}}(p,s).
\nonumber\\
\langle 0|\eta^{\Lambda }|\Lambda(p',s')\rangle &=&\lambda_{\Lambda} u_{\Lambda}(p',s'),
\nonumber\\
\langle 0|\eta_{\mu }^{J/\psi}|J/\psi(q)\rangle &=&f_{J/\psi} m_{J/\psi} \varepsilon_{\mu},
\label{eq:ResPc}
\end{eqnarray}
where $ \varepsilon_{\mu} $ is the polarization vector and  $ f_{J/\psi} $ is the decay constant of the $ J/\psi $ state, $\lambda_{P_{cs}}$, $\lambda_{\Lambda}$ are the current coupling constants of the $P_{cs}$ and $\Lambda$ states, $ u_{P_{cs}} $ and $ u_{\Lambda} $ are the corresponding spinors, respectively. $|P_{cs}(p,s)\rangle$ is used to represent one-particle pentaquark state with negative parity. The matrix element $\langle J/\psi(q) \Lambda(p',s')|P_{cs}(p,s)\rangle $ is given in terms of the coupling constants, $g_1$ and $g_2$ as
\begin{eqnarray}
\langle J/\psi(q) \Lambda(p',s')|P_{cs}(p,s)\rangle = \epsilon^{* \nu}\bar{u}_{\Lambda}(p',s')\big[g_1\gamma_{\nu}-\frac{i\sigma_{\nu\alpha}}{m_{\Lambda}+m_{P_{cs}}}q^{\alpha}g_2\big]\gamma_5 u_{P_{cs}}(p,s).
\label{eq:coupling}
\end{eqnarray}
In the next step, the matrix elements given in Eqs.~(\ref{eq:ResPc}) and (\ref{eq:coupling}) are placed in the Eq.~(\ref{eq:CorrF1Phys}) and following summations over spins of spinors and polarization vectors are applied
\begin{eqnarray}
\sum_{s}u_{P_{cs}}(p,s)\bar{u}_{P_{cs}}(p,s)&=&({\slashed
p}+m_{P_{cs}}),\nonumber \\
\sum_{s'}u_{\Lambda}(p',s')\bar{u}_{\Lambda}(p',s')&=&({\slashed
p'}+m_{\Lambda}), \nonumber\\
\varepsilon_{\alpha}\varepsilon^*_{\beta}&=&-g_{\alpha\beta}+\frac{q_\alpha q_\beta}{m_{J/\psi}^2}.
\label{eq:SumPc}
\end{eqnarray}
And finally, after the Borel transformation, which is applied to suppress the contributions coming from higher states and continuum, the result of physical side is obtained as
\begin{eqnarray}
\tilde{\Pi} _{\mu }^{\mathrm{Had}}(p, q)&=&e^{-\frac{m_{P_{cs}^2}}{M^2}}e^{-\frac{m_{\Lambda}^2}{M'^2}}\frac{f_{J/\psi} \lambda_{\Lambda} \lambda_{P_{cs}} m_\Lambda }{m_{J/\psi} (m_{\Lambda} + m_{P_{cs}}) (m_{J/\psi}^2 + Q^2)}\big[-g_1 (m_\Lambda + m_{P_{cs}})^2+g_2 m_{J/\psi}^2 \big]\not\!p p_{\mu}\gamma_5 \nonumber\\ 
&+&e^{-\frac{m_{P_{cs}^2}}{M^2}}e^{-\frac{m_{\Lambda}^2}{M'^2}}\frac{f_{J/\psi} \lambda_{\Lambda} \lambda_{P_{cs}} m_{J/\psi} m_\Lambda }{ (m_{\Lambda} + m_{P_{cs}}) (m_{J/\psi}^2 + Q^2)}\big[ g_1 (m_{\Lambda} + m_{P_{cs}})+g_2 (m_\Lambda - m_{P_{cs}}) \big] \not\!p \gamma_{\mu} \gamma_5 \nonumber\\
&+&\mathrm{other~structures}+\cdots
\label{eq:had}
\end{eqnarray}
where $M^2$ and $M'^2$ are the Borel parameters to be determined from the analyses imposing some necessary criteria and $Q^2=-q^2$. The result contains more Lorentz structures than the ones given explicitly in Eq.~(\ref{eq:had}). However, in the last equation, we present only the ones that are used directly in the analyses, and the others and the contribution of the excited states and continuum are represented as $\mathrm{other~structures}+\cdots$.  

The second representation of the correlation function, the QCD side, is obtained using the interpolating currents explicitly in the correlation function. To this end, the possible contractions between the quark fields are attained using Wick's theorem that renders the result to the one given in terms of heavy and light quark propagators as:
\begin{eqnarray}
\Pi_{\mu }^\mathrm{QCD}(p,p',q)&=&i^2\int d^{4}xe^{-ip\cdot x}\int d^{4}ye^{ip'\cdot y}\epsilon^{klm}\epsilon^{i'l'a'}\epsilon^{i'j'k'}\epsilon^{l'm'n'}\frac{1}{\sqrt{6}}\sum_{i=1}^2\bigg\{-2Tr[\gamma_{5}CS_{u}^{Tkj'}(y-x)C A_1^i S_{d}^{lk'}(y-x)]\nonumber\\
&\times& A_2^i S_s^{mm'}(y-x) \gamma_{5}C S_{c}^{Tnn'}(-x) C \gamma_{\mu} C S_{c}^{Ta'n}(x)C+A_2^iS_{d}^{mk'}(y-x)\gamma^{5}CS_{u}^{Tkj'}(y-x)C A_1^iS_{s}^{lm'}(y-x)\nonumber\\&\times& \gamma_{5}C S_{c}^{Tnn'}(-x) C \gamma_{\mu}C S_{c}^{Ta'n}(x)C+A_2^i S_{u}^{kj'}(y-x)\gamma^{5}CS_{d}^{mk'}(y-x)C A_1^iS_{s}^{lm'}(y-x) \gamma_{5}C \nonumber\\&\times& S_{c}^{Tnn'}(-x) C \gamma_{\mu}C S_{c}^{Ta'n}(x)C\bigg\},\label{eq:CorrF1Theore}
\end{eqnarray}
where $S_q^{ab}(x)=S_{u,d,s}^{ab}(x)$ and $S_{c}^{ab}(x)$ are the light and heavy quark propagators with the following explicit expressions:
\begin{eqnarray}
 S_{q}^{ab}(x)=&&i\frac{x\!\!\!/}{2\pi^{2}x^{4}}\delta_{ab}-\frac{m_{q}}{4\pi^{2}x^{2}}\delta_{ab}-\frac{\langle
 \overline{q}q\rangle}{12}\Big(1-i\frac{m_{q}}{4}x\!\!\!/\Big)\delta_{ab}-\frac{x^{2}}{192}m_{0}^{2}\langle
 \overline{q}q\rangle\Big( 1-i\frac{m_{q}}{6}x\!\!\!/\Big)\delta_{ab}\\
 &&
 \notag-\frac{ig_{s}G_{ab}^{\theta\eta}}{32\pi^{2}x^{2}}\Big[x\!\!\!/\sigma_{\theta\eta}+\sigma_{\theta\eta}x\!\!\!/ \Big]-\frac{x\!\!\!/ x^{2}g_s^2}{7776}\langle
 \overline{q}q\rangle^2\delta_{ab}+\cdots,
\end{eqnarray}
and
\begin{eqnarray}
\notag
 S_{c}^{ab}(x)=&&\frac{i}{(2\pi)^{4}}\int
 d^{4}ke^{-ik.x}\Big\{\frac{\delta_{ab}}{k\!\!\!/-m_{c}}-\frac{g_{s}G_{ab}^{\alpha\beta}}{4}\frac{\sigma_{\alpha\beta}(k\!\!\!/+m_{c})+(k\!\!\!/+m_{c})\sigma_{\alpha\beta}}{(k^{2}-m_{c}^{2})^{2}}\\
 &&+\frac{\pi^{2}}{3}\Big\langle\frac{\alpha_{s}GG}{\pi}\Big\rangle\delta_{ab}m_{c}\frac{k^{2}+m_{c}k\!\!\!/}{(k^{2}-m_{c}^{2})^{4}}+\cdots\Big\}.
\end{eqnarray}
The same Lorentz structures obtained in the hadronic side are also present in this side, and the ones used in our analyses are $\not\!p p_{\mu} \gamma_{5} $ and $ \not\!p \gamma_{\mu}\gamma_5 $, whose contributions are represented in the below equation explicitly, and the contributions of the others are represented with the last term stated as $\mathrm{other \,\,\, structures}$. 
\begin{eqnarray}
\Pi_{\mu }^{QCD}(p,q)&=&\Pi_1\,\not\!p p_{\mu} \gamma_{5} +
\Pi_2\,  \not\!p \gamma_{\mu}\gamma_5  +\mathrm{other \,\,\, structures}. \label{eq:PiOPE}
\end{eqnarray}
To obtain the coefficients, $\Pi_i$, of these Lorentz structures, we use the propagators explicitly in Eq.~(\ref{eq:CorrF1Theore}) and transform the results into momentum space. After computation of the four integrals the spectral densities, $\rho_i$ are obtained from the imaginary part of the results, $\rho_i(s,s',q^2)=\frac{1}{\pi}Im[\Pi_i]$. These spectral densities are used in the following dispersion relation:
\begin{eqnarray}
\Pi_{i}=\int ds\int
ds'\frac{\rho_{i}^{\mathrm{pert}}(s,s',q^{2})+\rho_{i}^{\mathrm{non-pert}}(s,s',q^{2})}{(s-p^{2})
(s'-p'^{2})}, \label{eq:Pispect}
\end{eqnarray}
giving us the final results of the QCD representation of the correlation function. In the last equation $i=1,2,..,12$ and $ \rho_{i}^{\mathrm{pert}}(s,s',q^{2}) $ and $ \rho_{i}^{\mathrm{non-pert}}(s,s',q^{2}) $ represent the perturbative and non-perturbative parts of the spectral densities, respectively. The results of the spectral densities that are used in the analyses ($ i=1,2 $) are:
\begin{eqnarray}
\rho_1^{\mathrm{pert}}&=&\int_0^1 dx \int_0^{1-x}dy \frac{1}{1024 \sqrt{
    6} \pi^6 \chi^3 \chi'^6}(1 + 5 \beta) m_s x y (Q^2 x y + s' \chi \chi' + 
   m_c^2 \chi'^2)^2\Theta[L(s,s',Q^2,x,y)],\nonumber\\
\rho_1^{\mathrm{non-pert}}&=& \int_0^1 dx \int_0^{1-x}dy \Bigg\{-\frac{1}{128 \sqrt{
   6} \pi^4 \chi^2 \chi'^5 }\big[\big((\beta -1) \langle\bar{d}d\rangle + \langle\bar{s}s\rangle(1 + 5 \beta) + (\beta -1 ) \langle\bar{u}u\rangle \big) x y \big(Q^2 x y + s' \chi \chi' + m_c^2 \chi'^2\big)\big]\nonumber\\
 &-&\frac{\langle \frac{\alpha_s G^2}{\pi}\rangle}{36864 \sqrt{6} \pi^4 \chi^4 \chi'^5} (1 + 5 \beta) m_c x^4 y^2
 +\frac{\langle \frac{\alpha_s G^2}{\pi}\rangle}{(9216 \sqrt{6} \pi^4 \chi^3 \chi'^4)} (1 + 5 \beta) m_s x y \big[9 x^2 + 9 (y-1 )^2 + x (19y-18 )\big]\nonumber\\
 &+&\frac{1}{256 \sqrt{6} \pi^4 \chi \chi'^4}\big[m_0^2 \big((\beta -1 )  \langle\bar{d}d\rangle + \langle\bar{s}s\rangle (1+ 5 \beta) + (\beta -1)  \langle\bar{u}u\rangle \big) x y\big]
 \Bigg\}\Theta[L(s,s',q^2,x,y)],
\end{eqnarray}
and 
\begin{eqnarray}
\rho_2^{\mathrm{pert}}&=&\int_0^1 dx \int_0^{1-x}dy \frac{1}{2048 \sqrt{6} \pi^6 \chi^3 \chi'^4}\Big[-(1 + 5 \beta) m_c m_s \big(Q^2 x y + m_c^2 \chi'^2 + s' (x^2 + (y-1) y + x (2y-1))\big)^2\Big]\nonumber\\
&\times& \Theta[L(s,s',Q^2,x,y)]\nonumber\\
\rho_2^{\mathrm{non-pert}}&=& \int_0^1 dx \int_0^{1-x}dy \Bigg\{\frac{1}{512 \sqrt{6} \pi^4 \chi^2 \chi'^4}\Big[m_s \big(\langle\bar{s}s\rangle(1  + 5 \beta) - 2 (\beta -1) \langle\bar{u}u\rangle \big) \chi \big(2 Q^2 x y + 
    s' \chi \chi'\big) - 
 2 m_c \big(\langle\bar{s}s\rangle( 1 + 5 \beta ) \nonumber\\
 &+& (\beta -1 ) \langle\bar{u}u\rangle \big) \chi' \big(Q^2 x y + 
s' \chi \chi' + m_c^2 \chi'^2 \big) - 
 2 (\beta -1) \langle\bar{d}d\rangle  \big(m_s \chi (2 Q^2 x y + 
       s' \chi \chi') + 
    m_c \chi' (Q^2 x y + s' \chi \chi' \nonumber\\
    &+& m_c^2 \chi'^2)\big)\big]+
 \frac{\langle \frac{\alpha_s G^2}{\pi}\rangle}{147456 \sqrt{6} \pi^4 \chi^4 \chi'^4}(1 + 5 \beta) m_c \big[m_c x y \chi' (4 x^2 - 3 x y + 4 y^2) - 
 8 m_s \chi (2 x^4 - x^3 y - x y^3 + 2 y^4)\big]\nonumber\\
 &-&\frac{\langle \frac{\alpha_s G^2}{\pi}\rangle}{36864 \sqrt{6} \pi^4 \chi^3 \chi'^4}(1 + 5 \beta)
 \big[-(Q^2 + s - s') x^2 y \chi + 
 2 m_c m_s \chi'^2 \big(9 x^2 + 9 (y-1 )^2 + x (17y-18 )\big)
\Big]\nonumber\\
&-&\frac{m_0^2}{1536 \sqrt{6} \pi^4 \chi \chi'^4} \big[  2 m_s \big(\langle\bar{s}s\rangle(1 + 5 \beta) - 3 (\beta -1 ) \langle\bar{u}u\rangle\big) x y + 
   3 m_c \big(\langle\bar{s}s\rangle(1 + 5 \beta ) + (\beta -1) \langle\bar{u}u\rangle \big) \chi'^2 \nonumber\\
   &+& 
   3 (\beta -1 ) \langle\bar{d}d\rangle \big(-2 m_s x y + m_c \chi'^2\big)\big]+\frac{1}{5184 \sqrt{6} \pi^4 \chi \chi'^4}(1 + 5 \beta) g_s^2 (\langle\bar{d}d\rangle ^2 + \langle\bar{s}s\rangle ^2 +\langle\bar{u}u\rangle ^2) x y\nonumber\\
   &+&\frac{1}{48 \sqrt{6} \pi^2 \chi \chi'^4}\big[(\beta -1 ) \langle\bar{s}s\rangle \langle\bar{u}u\rangle + \langle\bar{d}d\rangle \big((\beta-1 ) \langle\bar{s}s\rangle + \langle\bar{u}u\rangle + 5 \beta \langle\bar{u}u\rangle\big)\big] x y
 \Bigg\}\Theta[L(s,s',Q^2,x,y)]
\end{eqnarray}
where
\begin{eqnarray}
\chi&=&(x+y-1),\nonumber\\
\chi'&=&(x+y),\nonumber\\
L(s,s',Q^2,x,y)&=&\frac{-Q^2 x y - s' (x+y-1) (x + y) - m_c^2 (x + y)^2}{(x + y)^2},
\end{eqnarray}
with $\Theta[\ldots]$ being the unit-step function.

\begin{figure}[h!]
\begin{center}
\includegraphics[totalheight=5cm,width=7cm]{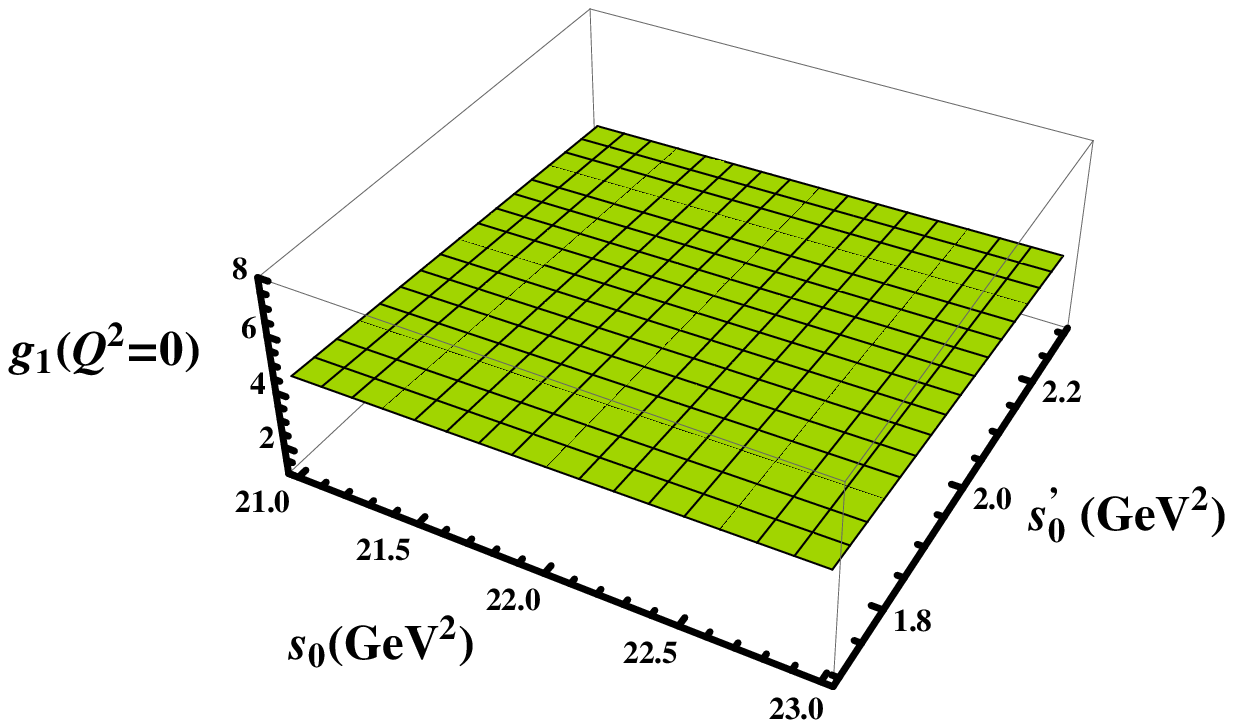}
\includegraphics[totalheight=5cm,width=7cm]{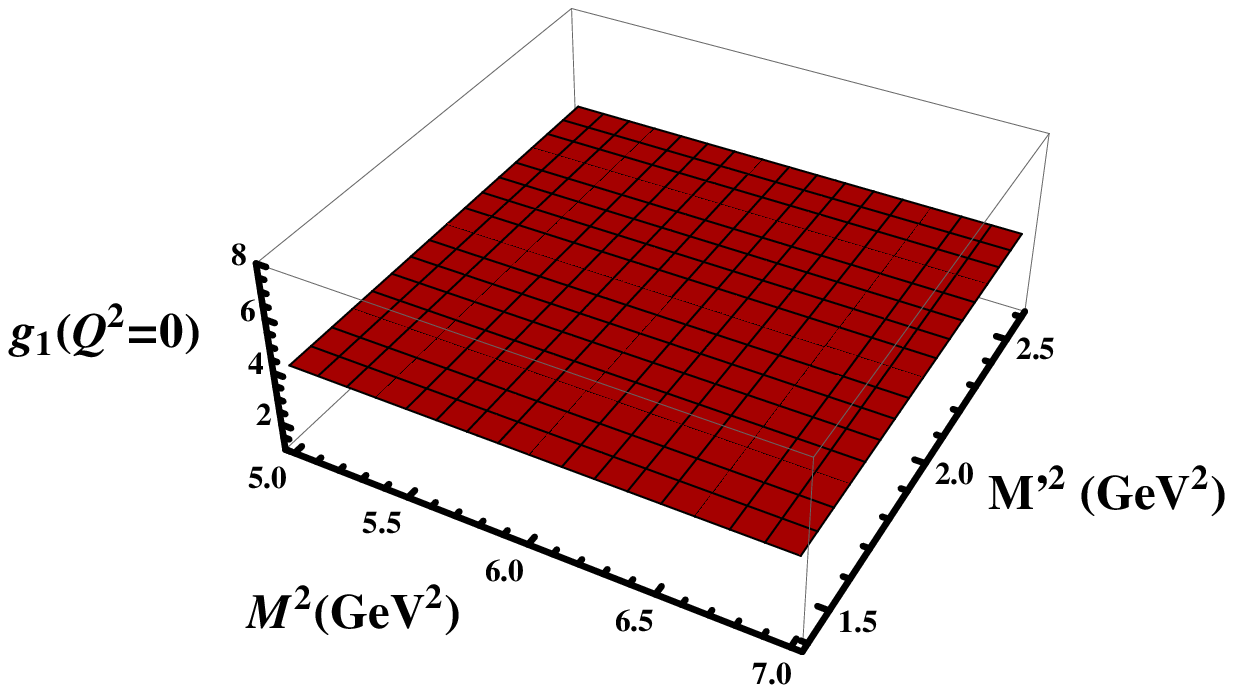}
\end{center}
\caption{\textbf{Left:} The variation of the strong coupling constant $g_1(Q^2=0)$ as a function of threshold parameters $s_0$ and $s_0'$ at the central values of the Borel parameters $M^2$ and $M'^2$ and the parameter $\beta$. \textbf{Right:} The variation of the strong coupling constant $g_1(Q^2=0)$ as a function of Borel parameters $M^2$ and   $M'^2$ at the central values of the threshold parameters $s_0$ and $s_0'$ and the parameter $\beta$.}
\label{gr:g1}
\end{figure}

\begin{figure}[h!]
\begin{center}
\includegraphics[totalheight=5cm,width=7cm]{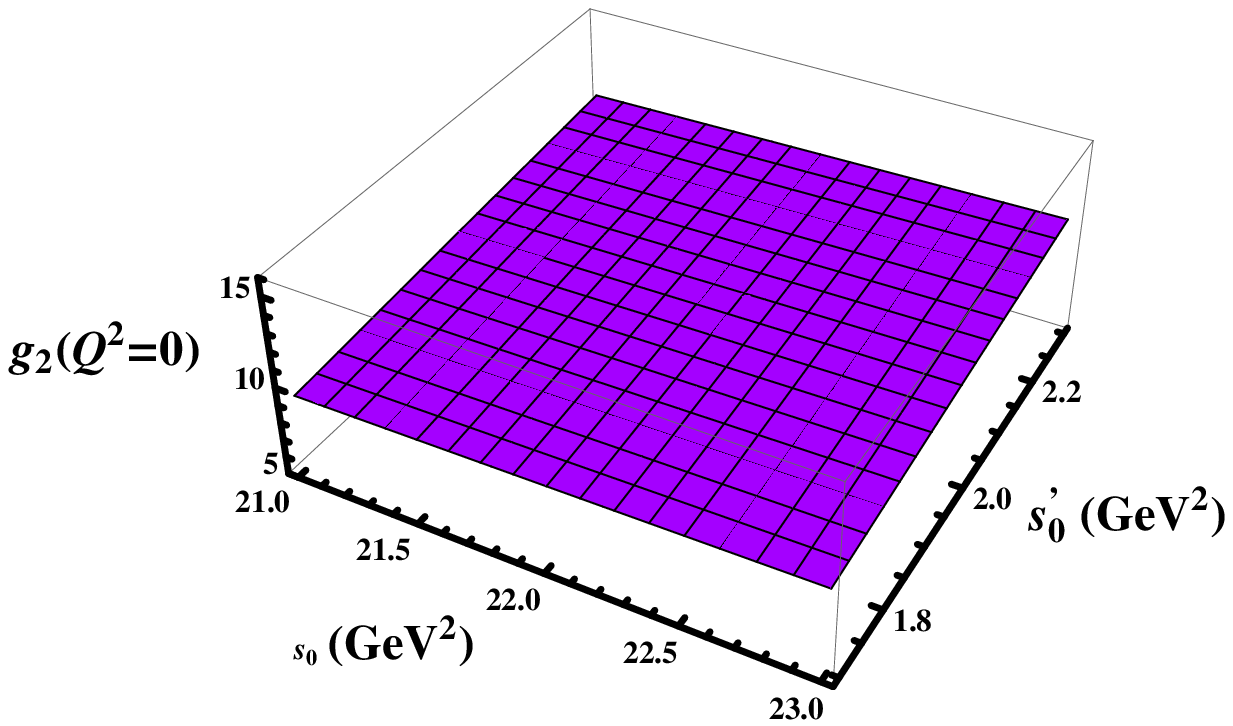}
\includegraphics[totalheight=5cm,width=7cm]{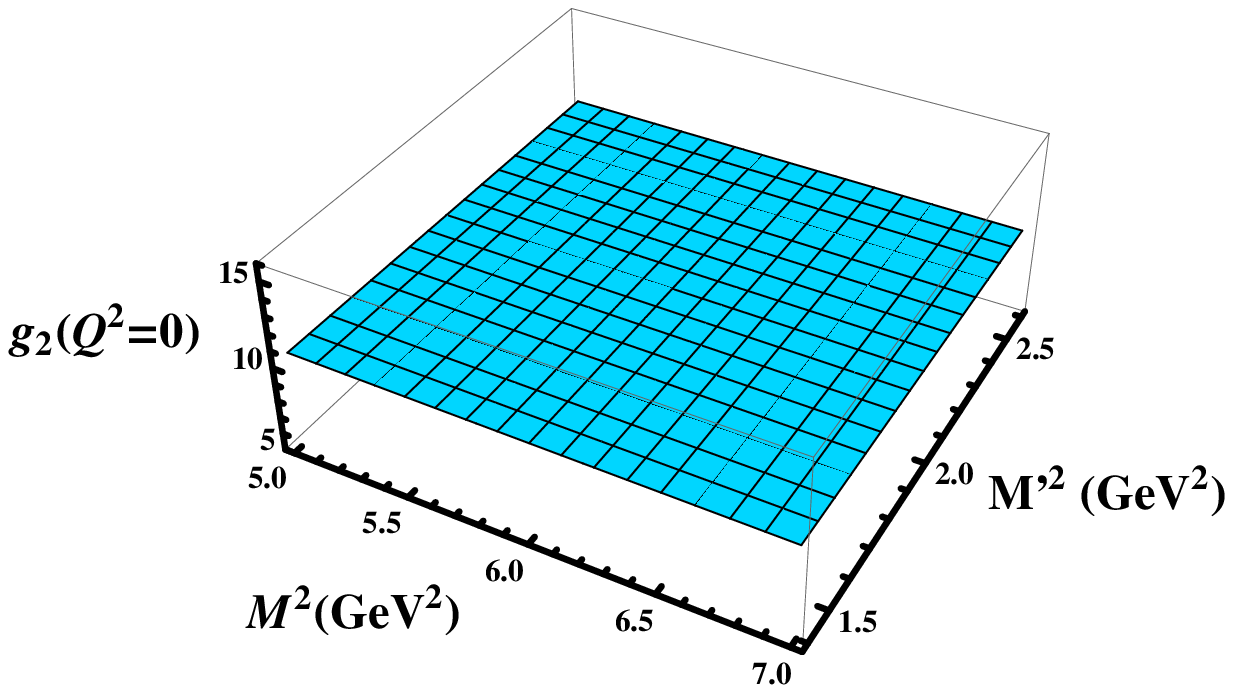}
\end{center}
\caption{\textbf{Left:} The variation of the strong coupling constant $g_2(Q^2=0)$ as a function of threshold parameters $s_0$ and $s_0'$ at the central values of the Borel parameters $M^2$ and $M'^2$ and the parameter $\beta$. \textbf{Right:} The variation of the strong coupling constant $g_2(Q^2=0)$ as a function of Borel parameters $M^2$ and   $M'^2$ at the central values of the threshold parameters $s_0$ and $s_0'$ and the parameter $\beta$.}
\label{gr:g2}
\end{figure}

Completing the calculations of both representations, we match the results considering the coefficients of the same Lorentz structures. This step gives two results both containing $g_1$ and $g_2$. Solving these two coupled expressions together, we obtain the sum rules giving the considered coupling constants, $g_1$ and $g_2$, analytically as
\begin{eqnarray}
g_1&=&
e^{\frac{m_{P_{cs}}^2}{M^2}}e^{\frac{m_\Lambda^2}{M'^2}}\frac{m_{J/\psi}(m_{J/\psi}^2+Q^2)
\left[(m_{P_{cs}}-m_{\Lambda})\tilde{\Pi}_1 +\tilde{\Pi}_2 \right]}
{f_{J/\psi}\lambda_{\Lambda}\lambda_{P_{cs}}m_{\Lambda}(m_{\Lambda}^2+m_{J/\psi}^2-m_{P_{cs}}^2)}\nonumber \\
g_2&=& e^{\frac{m_{P_{cs}}^2}{M^2}}e^{\frac{m_\Lambda^2}{M'^2}}\frac{(m_{P_{cs}}+m_{\Lambda})(m_{J/\psi}^2+Q^2)
\left[m_{J/\psi}^2\tilde{\Pi}_1+(m_{P_{cs}}+m_{\Lambda})\tilde{\Pi}_2  \right]}
{f_{J/\psi}\lambda_{\Lambda}\lambda_{P_{cs}}m_{\Lambda}m_{J/\psi}(m_{\Lambda}^2+m_{J/\psi}^2-m_{P_{cs}}^2)},
  \label{eq:g1}
\end{eqnarray}
where  $\tilde{\Pi}_i$ is the Borel transformed form of the $\Pi_i$ function. As is seen from the results for computations of these coupling constants, we are in need of some input parameters. These parameters are collected in Table~\ref{tab:Param}. In the calculations, the masses of light $u$ and $d$ quarks are taken as zero.
\begin{table}[tbp]
\begin{tabular}{|c|c|}
\hline\hline Parameters & Values \\ \hline\hline
$m_{s}$                                   & $93^{+11}_{-5}~\mathrm{MeV}$ \cite{Zyla:2020zbs}\\
$m_{c}$                                   & $(1.27\pm 0.02)~\mathrm{GeV}$ \cite{Zyla:2020zbs}\\
$m_{P_{cs}}$                              & $(4.47\pm0.11)~\mathrm{GeV}$ \cite{Wang:2020eep}\\
$m_{J/\psi}$                              & $(3096.900\pm0.006)~\mathrm{MeV}$ \cite{Zyla:2020zbs}\\
$m_{\Lambda}$                             & $( 1115.683\pm 0.006 )~\mathrm{MeV}$ \cite{Zyla:2020zbs}\\
$\lambda_{P_{cs}}$                        & $(1.86\pm 0.31)\times 10^{-3}~\mathrm{GeV}^6$ \cite{Wang:2020eep}\\
$\lambda_{\Lambda}$                       & $(0.013 \pm 0.02)~\mathrm{GeV}^3$ \cite{Aliev:2002ra}\\
$f_{J/\psi}$                              & $(481\pm36)~\mathrm{MeV}$ \cite{Veliev:2011kq}\\
$\langle \bar{q}q \rangle $               & $(-0.24\pm 0.01)^3$ $\mathrm{GeV}^3$~\cite{Belyaev:1982sa}  \\
$\langle \bar{s}s \rangle $               & $0.8\langle \bar{q}q \rangle$ \cite{Belyaev:1982sa} \\
$m_{0}^2 $                                & $(0.8\pm0.1)$ $\mathrm{GeV}^2$ \cite{Belyaev:1982sa}\\
$\langle \overline{q}g_s\sigma Gq\rangle$ & $m_{0}^2\langle \bar{q}q \rangle$ \\
$\langle g_s^2 G^2 \rangle $              & $4\pi^2 (0.012\pm0.004)$ $~\mathrm{GeV}
^4 $\cite{Belyaev:1982cd}\\
\hline\hline
\end{tabular}%
\caption{Some input parameters entering the calculations.}
\label{tab:Param}
\end{table}   
Besides the given parameters in Table~\ref{tab:Param}, there are five additional parameters which are the threshold parameters $s_0$ and $s'_0$, Borel parameters, $M^2$ and $M'^2$ and the mixing parameter $\beta$ which is coming from the interpolating current of the $\Lambda$ baryon. These parameters are determined from the analyses of the results imposing the standard criteria of the method such as weak dependence of the results on the auxiliary parameters, pole dominance and  convergence of operator product expansion (OPE).  Considering these conditions, the threshold parameters are fixed as follows:
\begin{eqnarray}
21.0\,\,\mathrm{GeV}^{2}&\leq& s_{0}\leq 23.0\,\,\mathrm{GeV}^{2},
\nonumber \\
1.7\,\,\mathrm{GeV}^{2}&\leq& s'_{0}\leq 2.3\,\,\mathrm{GeV}^{2}.
\label{Eq:s0s0p}
\end{eqnarray}
The  upper limits of Borel parameters are determined  by imposing the condition of pole dominance for the selected working regions of continuum thresholds. To this end, we consider the following ratio using the  continuum subtracted and Borel transformed invariant amplitude $\Pi_i(s_0,s'_0,M^2,M'^2,\beta)$ obtained from the QCD side:
\begin{eqnarray}
PC=\frac{\Pi_i(s_0,s'_0,M^2,M'^2,\beta)}{\Pi_i(\infty,\infty,M^2,M'^2,\beta)},
\end{eqnarray}
where $ PC $ denotes the pole contribution and $ i $ stands for the selected structures. To fix the upper limit of the Borel parameters we impose this ratio to be larger or at least equal to $20\%$, which is typical in the analyses of the exotic states.   For the calculations of their lower limits, the convergence of the series of OPE is considered: the dominance of perturbative part  over the nonperturbative ones and  "the higher the dimension of the nonperturbative operator, the lower its contribution". To extract the lower limit, using this criteria, we fix the ratio of the higher dimensional term, that is the term having dimension 6 in the QCD side, to the whole result as follows:
\begin{eqnarray}
R(M^2,M'^2)=\frac{\Pi_i^6(s_0,s'_0,M^2,M'^2,\beta)}{\Pi_i(s_0,s'_0,M^2,M'^2,\beta)},
\end{eqnarray}
and keep this ratio as $R(M_{min}^2,M'{}^2_{min})=0.02$ to certify the convergence of the OPE. 
With these conditions, the Borel parameters are fixed as 
\begin{eqnarray}
5.0\ \mathrm{GeV}^{2}\leq M^{2}& \leq & 7.0\ \mathrm{GeV}^{2},
\nonumber \\
1.4\ \mathrm{GeV}^{2}\leq M'^{2}& \leq & 2.6\ \mathrm{GeV}^{2}.
\label{Eq:MsqMpsq}
\end{eqnarray}%
As the final parameter, we determine the working intervals of $\beta$ from the analyses by considering a parametric plot of the results as  functions of  $\cos\theta$ where $\beta=\tan\theta$. We select the regions that show least variations with respect to the changes in $\cos\theta$, which read
\begin{eqnarray}
-1\leq\cos\theta\leq -0.5 ~~~~~\mbox{and} ~~~~~~0.5\leq \cos\theta\leq 1. 
\end{eqnarray}  
Our analyses show that the physical quantities show weak dependence on the auxiliary parameters in the above windows for $s_0$ and $s'_0$,  $M^2$, $M'^2$ and  $\cos\theta$. To depict the dependence of the results of coupling constants on the auxiliary parameters, we plot the Figures~\ref{gr:g1} and \ref{gr:g2} for $g_1$ and $g_2$ at $Q^2=0$. From these figures and the numerical values we see a good stability of the results with respect to  the Borel parameters in their working window. However, the results show some weak dependencies on the continuum thresholds in their working intervals, which remain inside the limits allowed by the method. The variations with respect to the auxiliary parameters appear as the main sources of the  uncertainties in the numerical results.

Using the given input parameters in Table~\ref{tab:Param} and the determined windows for auxiliary parameters, we calculate the strong  coupling constants for the considered decay channel. The following fit functions represent the  $Q^2$-behavior of the strong coupling form factors:
\begin{eqnarray}
g_i(Q^2)&=& g_{0}e^{c_1\frac{Q^2}{m_{P_{cs}}^2}+c_2(\frac{Q^2}{m_{P_{cs}}^2})^2}.
\end{eqnarray}
with $g_0$, $c_1$ and $c_2$ being the fit parameters that take the values given in Table~\ref{tab:FitParam}.
\begin{table}[tbp]
\begin{tabular}{|c|c|c|c|}
\hline\hline Coupling Constant  &$ g_0$ & $c_1$ & $c_2$ \\ \hline\hline
 $g_1$&$4.22 \pm 0.51$ &$1.54$ & $1.16$ \\
 \hline
   $g_2$&$10.54 \pm 1.26$ & $1.54$ & $1.16$ \\
\hline\hline
\end{tabular}%
\caption{: Parameters of the fit functions for
coupling constants, $g_1$ and $g_2$.} \label{tab:FitParam}
\end{table}
We, then, use the fit functions to determine the coupling constants at $Q^2=-m_{J/\psi}^2$ as
\begin{eqnarray}
g_1=2.63 \pm 0.31~~~~~~~~\mathrm{and}~~~~~~~~g_2=5.25 \pm 0.63,
\end{eqnarray}
where the errors are due to the uncertainties present in the input parameters entering the calculation and in the determinations of the auxiliary parameters, as well. 

Having determined the strong coupling constants, the next task is to compute  the corresponding width for $P_{cs}\rightarrow J/\psi \Lambda$ decay channel in terms of the strong coupling constants and other related parameters. The standard calculations lead to the width formula as  
\begin{eqnarray}
\Gamma &=& \frac{f(m_{P_{cs}},m_{J/\psi},m_{\Lambda}) }{16\pi m_{P_{cs}}^2}\Bigg[-\frac{2 (m_{J/\psi}^2 - (m_\Lambda + m_{P_{cs}})^2)}{m_{J/\psi}^2 (m_\Lambda + m_{P_{cs}})^2}\Big(g_2^2 m_{J/\psi}^2 (m_{J/\psi}^2 + 2 (m_\Lambda - m_{P_{cs}})^2) 
\nonumber\\
&+& 
6 g_1 g_2 m_{J/\psi}^2 (m_\Lambda - m_{P_{cs}}) (m_\Lambda + m_{P_{cs}}) + 
g_1^2 (2 m_{J/\psi}^2 + (m_\Lambda - m_{P_{cs}})^2) (m_\Lambda + m_{P_{cs}})^2 \Big)\Bigg],
\label{Eq:DW}
\end{eqnarray}
where
\begin{eqnarray}
f(x,y,z)&=&\frac{1}{2x}\sqrt{x^4+y^4+z^4-2x^2y^2-2x^2z^2-2y^2z^2}.
\end{eqnarray}   
Using the values of the  strong coupling constants, we compute the width for the considered channel to be
\begin{eqnarray}
\Gamma(P_{cs} \rightarrow J/\psi \Lambda)&=&\left(15.87\pm 3.11\right)~\mathrm{MeV}.
 \label{Eq:DWNegativeParity}
\end{eqnarray}

\section{Summary and conclusion}\label{Sum} 

The recently observed pentaquark state, the hidden-charmed strange $P_{cs}(4459)^0$, added a new member to the pentaquark family. Its experimentally observed mass and width were reported as $M=4458.8 \pm 2.9^{+4.7}_{-1.1}~\mathrm{MeV}$ and $\Gamma = 17.3 \pm 6.5^{+8.0}_{-5.7}~\mathrm{MeV}$, respectively~\cite{Aaij:2020gdg}. However, its quantum numbers, $J^P$, could not be 
determined as a result of insufficient statistics in the experiment~\cite{Aaij:2020gdg}. Using the QCD sum rule method, $P_{cs}(4459)^0$ state was studied both in the molecular form assigning its quantum numbers as $J^P=\frac{1}{2}^-$ or $\frac{3}{2}^-$~\cite{Chen:2020uif} and in the diquark-diquark-antiquark form with quantum numbers $\frac{1}{2}^-$~\cite{Wang:2020eep}.  Its mass was obtained  in these studies and compared to experimental data to shed light on its nature. Both of these interpretations resulted in mass predictions consistent with the experimental data creating a need for further investigations of this state, for instance  its width.

In this study, we investigated the strong $P_{cs} \rightarrow J/\psi \Lambda $ decay and obtained the strong coupling constants representing the amplitude of this decay using the QCD sum rule method. To this end,  we adopted  an interpolating current in the diquark-diquark-antiquark form for the substructure of this particle. In the analysis, we considered the quantum numbers of  $P_{cs}(4459)^0$ state as $J^P=\frac{1}{2}^-$. The obtained strong coupling constants were used in the determination of the corresponding width, which is obtained as $\Gamma(P_{cs} \rightarrow J/\psi \Lambda)=\left(15.87\pm 3.11\right)~\mathrm{MeV}$. Compared to the experimental value, the obtained width is in good consistency with experimental data, which  favors  the quantum numbers $J^P=\frac{1}{2}^-$ and compact pentaquark nature of diquark-diquark-antiquark form for $P_{cs}(4459)^0$ state. 
 
 In Ref.~\cite{Wang:2021itn}, also,  the authors have considered the molecular interpretations for this state and concluded that  it is either $\bar{D}\Xi'_c$ with $J^P=\frac{1}{2}^-$ and $I=0$ or $\bar{D}\Xi^*_c$ with $J^P=\frac{3}{2}^-$ and $I=0$. In Ref.~\cite{Zhu:2021lhd} $P_{cs}(4459)$ state was interpreted as $\Xi_c\bar{D}^*$ with $J^P=\frac{3}{2}^-$ without excluding the possibility of its being two-pole structure $\Xi_c\bar{D}^*$ states with $J^P=\frac{1}{2}^-$  and $J^P=\frac{3}{2}^-$. Another molecular interpretation was given in Ref. \cite{Chen:2021tip} in which its two-body strong decay behaviors supported its being $\Xi_c\bar{D}^*$ state with $I(J^P)=0(\frac{3}{2}^-)$. 
 

 All of the above-mentioned  investigations indicate that the quantum numbers and nature of the  $P_{cs}(4459)^0$ state are still ambiguous and need clarification not only from further theoretical studies of its various properties but also from future experiments. Comparison of the theoretical results on various parameters of this state with future experimental data will shed light on the nature, quark-gluon organization and quantum numbers of this state.

\section*{ACKNOWLEDGEMENTS}
K. Azizi is thankful to Iran Science Elites Federation (Saramadan)
for the partial  financial support provided under the grant number ISEF/M/99171.



\end{document}